\documentstyle[preprint,aps]{revtex}

\begin{document}

\draft

\title{ Quantum Evaporation  from the Free Surface \\ of 
Superfluid $^4$He}     

\author{F. Dalfovo$^a$, A. Fracchetti$^a$, A. Lastri$^{a,b}$,  
L. Pitaevskii$^{c,d}$ and  S. Stringari$^a$}

\address{$^a$ Dipartimento di Fisica, Universit\`a di Trento, \\
    and Istituto Nazionale di Fisica della Materia, I-38050 Povo, Italy } 
\address{$^b$ Institut f\"ur Theoretische Physik, Universit\"at zu 
    K\"oln, D-50937 K\"oln, Germany}
\address{$^c$ Department of Physics, TECHNION, Haifa 32000,  Israel}
\address{$^d$ Kapitza Institute for Physical Problems, ul. Kosygina 2, 
    117334 Moscow }

\date{ March 8, 1996 }
\maketitle

\begin{abstract}
The scattering of atoms and rotons at the free surface of superfluid $^4$He 
is  studied in the framework of linearised time dependent mean field 
theory. The phenomenological Orsay-Trento density  functional is used to
solve  numerically the equations of motion for the elementary excitations
in presence of a free surface and to calculate the flux of rotons  and atoms
in the reflection, condensation, and evaporation processes.   The 
probability associated with each process  is evaluated as a function of 
energy,  for incident angles such that only rotons and atoms are
involved in the scattering (phonon forbidden region). The  evaporation
probability for $R^+$ rotons (positive group velocity) is  predicted to
increase quite rapidly from zero, near the roton minimum, to  $1$ as the
energy increases. Conversely the evaporation from $R^-$ rotons (negative 
group velocity) remains smaller than 0.25 for all energies. Close to the 
energy of the roton minimum, $\Delta$, the mode-change process $R^+ 
\leftrightarrow R^-$ is  the dominant one. The consistency  of the results
with general properties of the scattering matrix,  such as
unitarity and time reversal, is explicitly discussed.  The condensation of
atoms  into bulk excitations is also investigated. The condensation
probability is  almost $1$ at high energy in agreement with experiments, but
it lowers  significantly  when the energy approaches  the roton minimum in
the phonon forbidden region. 
\end{abstract}


\narrowtext

\section{Introduction}
\label{sec:intro}

One of the unique features of superfluid helium at low temperature is that
elementary  excitations like rotons and high energy phonons have a very  
long mean  free path, long enough to propagate ballistically in the liquid
on  macroscopic distances. Furthermore their energy is comparable with the 
energy required to eject atoms from liquid to vacuum. Thus rotons and 
phonons impinging on the free surface of the liquid can evaporate atoms 
by  one-to-one quantum processes.  One can also produce collimated 
beams of elementary excitations and  collect the evaporated atoms above 
the surface in order to  extract information on the properties of the 
superfluid. Similarly   one can investigate the opposite process of atoms 
which condense in the liquid by producing elementary excitations.  Several 
experiments have been carried out in the last  
decades\cite{Joh66,Bal78,Edw82,Bai83,Wya84,Wya90,Bad94,Ens94,Wya95,For95} 
to explore the phenomenon of quantum evaporation and condensation 
(see also Ref.~\cite{Wya91} for a recent review). From the theoretical 
viewpoint\cite{Wid69,And69,Col72,Car76,Clo92,Mar92,Mul92,Dal95}
 these processes are very appealing, being clean examples of 
scattering  of excitations in a highly correlated many-body quantum system.  
A  good theory of quantum evaporation and condensation is however a 
difficult task and a clear understanding of the fundamental mechanism 
underlying these phenomena has still to come, even though relevant 
steps in this direction have been accomplished in the last years.   

In this paper we present a time dependent mean field theory, based on the 
density functional formalism, which allows one to calculate the scattering 
matrix elements associated with the scattering of rotons, phonons  and atoms
at the free  surface.  Preliminary results of this theory have been recently
published in  Ref.~\cite{Dal95}. Here we discuss the theory in more detail
and we present  results in a wider range of energy and wave vectors. 

A discussion about quantum evaporation and condensation requires first a 
detailed knowledge of the spectrum of elementary excitations of superfluid 
$^4$He.  A schematic picture is given in Fig.~\ref{fig:spectrum}. The 
liquid-vacuum interface is supposed to be in the $xy$-plane. The system is 
translational invariant in those directions, so that the parallel wave vector 
$q_x$ is a conserved quantity  (we choose $q_y=0$ without any loss of 
generality).  The minimum energy needed to produce free atom 
states outside the liquid at zero temperature  is the chemical potential  
$|\mu|=7.15$ K.  The  threshold for evaporation for given values of $q_x$ 
is then $|\mu|+ \hbar^2 q_x^2/2m$ (dashed line). Phonons, rotons and atoms
can  propagate at different angles with respect to the surface, that is with 
different $q_z$ for a given $q_x$. The  dispersion law for phonons and
rotons propagating in the bulk liquid and in the direction parallel to the
surface ($q_z=0$) is plotted  as a function of $q_x$. Excitations
having $q_z \ne 0$ fill different regions of the  spectrum
in Fig.~\ref{fig:spectrum} with a continuum of states. For  instance
each point in region V corresponds to a roton $R^+$ (positive  group
velocity) propagating with  $q=\sqrt{q_x^2+q_z^2}$  such that
$\hbar  \omega (q)$ is the $R^+$ roton dispersion. Since $q_x \le q$, roton
states with $q_z \ne 0$ fill the  region at the left of the roton branch
plotted in the figure. Similarly in  region IV  also $R^-$ rotons (negative
group velocity) can propagate; each point in that region can be a
combination of $R^+$ and $R^-$ rotons  propagating at different angles.
Above the threshold for evaporation, in  region III, atom states outside the
liquid are possible too. Similarly in  region II  one has phonons, rotons as
well as atoms; in region I, below the energy  of the roton minimum,
$\Delta$, there are only phonons and atoms;  finally  in region VI, above
the maxon energy, there are rotons and atoms.   The free surface acts as a
scattering region for all these excitations. 

In Refs.~\cite{Mar92,Mul92} the effect of the surface on the propagation of 
the elementary excitations has been studied within a local density 
approximation. The dispersion law of the excitations was assumed to be the 
same as for a uniform liquid of density equal to the local density.  
Maris\cite{Mar92} used a simple  interpolation between the phonon-roton
dispersion law in bulk liquid and  the dispersion law of the free atoms. He
studied the behavior of classical  trajectories for quasiparticles along the
$z$-direction assuming the  conservation of energy and parallel momentum. 
From that analysis one  finds some interesting   constraints on the
structure of the classical  trajectories crossing  the free surface.  For
instance only phonons and rotons above the maxon energy  (about $14$ K) are
found to  evaporate  atoms.  Vice-versa, the theory predicts no evaporation
from rotons with   energy smaller than the maxon  energy, because of the 
occurrence  of a  barrier at the interface. The experimental evidence
\cite{Wya91} for    quantum evaporation  induced by  rotons even below the
maxon  energy   is  consequently  an  important indication of the crucial
role played by quantum  effects.  Quantum effects were partially included in
the calculations of  Ref.~\cite{Mul92} by means of perturbation theory based
on WKB states; the  density profile in that case was approximated by a 
suitable analytic  function and the dispersion law for the excited states, in
local density  approximation,  was calculated using Beliaev's theory with an
effective  interaction.  

For an  accurate treatment of  quantum effects one has to go beyond the 
local density  approximation and the semiclassical treatment. To 
accomplish the task we  proceed in two distinct steps: 

\begin{itemize}

\item We first evaluate the ground state and the excited states of the 
inhomogeneous system within a self-consistent theory sufficiently accurate to
reproduce  the main features of the spectrum in Fig.~\ref{fig:spectrum},
including the  structure and the excitations of the free surface. 

\item We identify the current carried by the elementary
excitations  and   use the numerical solutions of the equations of motion to
calculate  the asymptotic flux of excitations associated with a given
scattering  process. While this identification is trivial in the description
of free atoms in vacuum, it becomes less obvious in the liquid where
many-body effects, present in this highly correlated system, must be
properly included.   Taking the ratios of incoming and
outgoing fluxes one finally gets  the evaporation and condensation
probabilities, which are the final goal of  the theory. 

\end{itemize}

For the first step we use a time dependent mean field theory  based on the 
least action principle applied to an energy functional  $\int\!d{\bf r} 
{\cal  H} [\Psi^*,\Psi]$.  For the form  of the functional we follow
the recent proposal of  Ref.~\cite{Dal95b}.  We consider linear  variation
of   $\Psi$, in the form  $(\Psi_0+\delta \Psi)$, and expand  $\delta \Psi$ 
in plane waves along the parallel direction.   We restrict  ourselves to
variations of the energy linear in $\delta \Psi$. For this  reason the
theory describes  only  one-to-one processes (evaporation of one  atom by
one roton, creation of one roton by condensing one atom, etc.) and  not 
processes involving more than two excitations (for example  multi-phonon or
multi-ripplon production in the atom condensation process).   It is worth
mentioning  that some experiments by Wyatt and 
co-workers\cite{Wya84,Wya90,Wya91}   support the idea  that the evaporation 
of atoms from bulk excitations is essentially  a one-to-one process, so
that  the use of a linear theory  seems 
justified  in this case. Conversely  measurements of atom condensation 
\cite{Edw82,Wya95}  point out an important role played by non linear
effects  associated  with the creation  of ripplons. This 
asymmetry  between  evaporation and  condensation  is still basically
unexplained.   Quantitative predictions for the scattering rates in  linear
theory are essential in order to better understand  the origin of this and
other discrepancies. 

As concerns the second step we derive the definition of  flux (or current 
density)  associated with the propagation of the elementary excitations, in 
terms of the linearised solutions of the equations of motion. We show  
that   such a current density obeys the proper equation of continuity.  We
will  devote detailed discussions about the meaning of the flux of
excitations  and about the properties of the scattering matrix elements,
which are  calculated starting from the knowledge of the fluxes involved in 
the  scattering process.  

The theory can be applied  to all the relevant regions of the spectrum in 
Fig.~\ref{fig:spectrum}.  In the present work we present results for 
processes involving rotons and atoms in the region III.  We chose this 
region for two basic reasons: first, phonons are not allowed there and thus 
the analysis is simpler;  second, the arguments based on classical 
trajectories predict no evaporation in this region and this means that
quantum effects should play a crucial role in the evaporation process. 
The analysis of the  remaining parts of the spectrum will be the object of
a future work. 

The paper is organised as follows: in Sec.~\ref{sec:2} we present the 
theoretical formalism of the density functional and the equations of motion. 
In Sec.~\ref{sec:3} we introduce the concept of current of excitations. In 
Sec.~\ref{sec:4} we discuss general properties of the scattering matrix. 
In  Sec.~\ref{sec:5}  we present the results for the probabilities of 
evaporation and  condensation.   Finally we will give a 
brief  summary of the main results and discuss some future perspectives. 

\section{Time dependent density functional theory}
\label{sec:2} 

The formalism used in the present work was already introduced in 
Ref.~\cite{Dal95b}. In that paper detailed discussions were devoted to the 
motivations of the theory and to the choice of the density functional. 
Results  for static and dynamic properties of superfluid $^4$He in different 
geometries were presented.  In Refs.~\cite{Las95,Cas95} the dynamics of the 
free surface and of droplets, respectively, were studied in detail within the 
same scheme. As a consequence we refer to  the above-mentioned papers for
details and  we recall here simply the main  points. 

The starting assumption is that the energy  can be written as  a functional 
of the form 
\begin{equation}
     E \  = \   \int \!d{\bf r} \ {\cal H} [\Psi, \Psi^*] \ \ \ 
\label{eq:e}
\end{equation}
where the complex function $\Psi$ is written as 
\begin{equation}
\Psi ({\bf r},t) = \Phi({\bf r},t) \exp \left({i \over \hbar} S ({\bf r},t) 
\right)  \ \ \ .
\label{eq:psirt}
\end{equation}
The real function $\Phi$ is related to the diagonal one-body density by 
$\rho=\Phi^2$. The phase $S$ fixes the velocity of the 
fluid  through the relation  ${\bf v} = (1/m) \nabla S$, where $m$ is the 
mass of the $^4$He atoms.  In the  calculation of the ground state  only 
states with zero velocity  must be considered, so that the energy is a
functional  only of the particle density $\rho({\bf r})$. A natural
representation is  given 
by 
\begin{equation}
E \ = \ \int \! d{\bf r} \ {\cal H}_0 [\rho] \ =  \  E_c[\rho] \ + \  
\int \! d{\bf r}  \ {\hbar^2 \over 2m}  ( \nabla \sqrt{\rho})^2  \ \ \ ,
\label{eq:ec}
\end{equation}
where the second term on the r.h.s. is a quantum pressure, corresponding 
to  the kinetic  energy of a Bose gas of non uniform density. The quantity 
$E_c[\rho]$ is  a {\sl correlation energy}; it incorporates the effects of 
dynamic correlations induced by the interaction.   Ground state 
configurations are obtained by minimising the energy of 
the system with respect to the density. This leads to the Hartree-type
equation 
\begin{equation}                                                     
( H_0 - \mu )   \sqrt{\rho({\bf r})} \ = \ 0 \ \ \ ,
\label{eq:hartree}
\end{equation}
where
\begin{equation}                                                     
H_0 = - {\hbar^2 \over 2m} \nabla^2  + U [\rho, {\bf r}] 
\label{eq:h0}
\end{equation}
is an effective Hamiltonian. The quantity $U[\rho, {\bf r}] \equiv \delta 
E_c 
/\delta \rho({\bf r})$ acts as a  mean field,  while the chemical potential 
$\mu$ is introduced in order to ensure the  proper normalisation  of the 
density to a fixed number of  particles. The dynamics of the system can be 
studied by using the least action principle:  
\begin{equation}
\delta \int_{t_1}^{t_2} dt \int \! d{\bf r} \left[ {\cal H} [\Psi^*, \Psi]
-\mu \Psi^* \Psi - \Psi^* i \hbar {\partial \Psi \over \partial t } 
\right] \ =\ 0 
\ \ \ .
\label{eq:leastaction}
\end{equation}
By  making variations with respect to $\Psi$ or $\Psi^*$ one derives the  
equations of motion for the excited states, in the form
of Schr\"odinger-like  equations. The equation for $\Psi$ is 
\begin{equation}
	(H - \mu )  \Psi = i \hbar {\partial \over \partial t } \Psi \ \ \ ,
\label{eq:schro}
\end{equation}
where $H = \delta E/ \delta \Psi^*$ is an effective Hamiltonian.  We 
linearise the equation by writing 
\begin{equation}
	\Psi ({\bf r},t) = \Psi_0({\bf r}) + \delta \Psi({\bf r},t)
\label{eq:linearised}
\end{equation}
where $\Psi_0({\bf r})$ corresponds to the ground state. The change of
wave function $\delta \Psi$ can be written in the form 
\begin{equation} 
\delta \Psi ({\bf r},t) = f({\bf r}) e^{-i \omega t} +
g^*({\bf r}) e^{i \omega t} \; , 
\label{eq:deltapsi1}
\end{equation}
where the functions $f({\bf r})$ and $g({\bf r})$ are fixed, 
together with the frequency $\omega$, by the solution of the equations of
motion (\ref{eq:schro}).  The Hamiltonian  $H$
then takes the form  
\begin{equation}
	H = H_0 + \delta H 
\label{eq:deltah}
\end{equation}
and the equation of motion becomes
\begin{equation}
(H_0 - \mu ) \delta \Psi + \delta H \ \Psi_0 \ = \ i \hbar {\partial \over
\partial t } \delta \Psi \ \ \ . 
\label{eq:schrolin}
\end{equation}
The term  $\delta H$ is linear in $\delta \Psi$ and accounts 
for changes in the Hamiltonian induced by the collective
motion of the system. Since $H$ depends explicitly
on the wave function $\Psi$,  Equation (\ref{eq:schro}) has to be solved
using a self-consistent procedure. 

The above scheme corresponds to a time dependent density functional 
(TDDF) theory. A discussion about equivalent formulations of the same 
approach (hydrodynamic-like equations, Green's functions), as well as about 
the connection with other many body approaches, is given in 
Refs.~\cite{Dal95b,Las95,Cas95}. Equations of the same type are obtained
using Correlated Basis Functions in the context of Hypernetted Chain
approximation \cite{Cam}.  Here we stress that the theory corresponds  to  a
quantum mechanical treatment of the dynamics and  consequently   accounts
for  the interference and tunneling phenomena  which are  expected  to play a
crucial role in the evaporation process. Of course, due to   linearisation,
they do not include inelastic  processes  associated with  multi-phonons or 
multi-ripplons. As already said in the Introduction,  these effects lie
beyond  the present theory.

To proceed further one has to specify both the explicit form of  the  
energy functional  (\ref{eq:e}) and the geometry of the system under 
investigation. The energy functional is the one introduced in 
Ref~\cite{Dal95b}. It consists of two parts, a functional
of the particle  density only and a functional of both the particle
density and the  current density. It corresponds to a generalisation of a
previous finite-range  functional  introduced by Dupont-Roc {\it et
al.}\cite{Dup90}. Its explicit form is given  in Appendix A. The functional
is phenomenological, i.e., it contains  parameters which are fixed to
reproduce known properties of the bulk  liquid. In particular the equation
of state and  the static response function  in bulk liquid are reproduced by
construction. The new functional, with  respect to the one of
Ref.~\cite{Dup90}, contains a current-current effective  interaction, which
accounts for backflow-like correlations and allows one  to reproduce the
experimental phonon-roton spectrum in the uniform  liquid. The latter is a
crucial requirement for a theory whose goal is the  prediction of
evaporation and condensation probabilities, since these  quantities depend
dramatically on the form of the excitation spectrum.

As concerns the geometry, we assume translational invariance in the 
$x$-$y$ plane. The density profile of the liquid in the ground state is a 
function  of the orthogonal co-ordinate $z$.  The profile of the free surface
obtained  with the present density functional theory is about $6$ \AA\  thick
and  compares  well with the one coming from ab initio calculations as
well as with the experimental data of Ref.~\cite{Lur92} (see 
Ref.~\cite{Cas95} for a discussion). The excited states are described by 
fluctuations of the particle density and velocity associated with the
solutions (\ref{eq:deltapsi1}) of the equations of motion. In the planar
geometry the same solutions can be written in the form\cite{note1}
\begin{equation}   
\delta \Psi({\bf r},t) \  = \ f(z) e^{i (q_x x -  \omega
t) } +  g(z) e^{- i (q_x x - \omega t) }  \; . 
\label{eq:deltapsi}
\end{equation}
With a proper choice for the boundary conditions the functions $f(z)$ and
$g(z)$ can be chosen real. Once the equations
of motion are written explicitly for the  unknowns $f(z)$ and $g(z)$, they
assume the typical form of the equations  of the random phase approximation
(RPA) for bosons. In particular they account for  both  particle-hole
[$f(z)$]  and hole-particle [$g(z)$] transitions which  are coupled by the
equations of motion  (\ref{eq:leastaction}). This   coupling is of crucial
importance in order  to treat  the  correlation effects associated with the
propagation of  elementary  excitations in an interacting system. The
equations of motion have also a structure formally  identical to the one  of
the Bogoliubov  equations for the dilute Bose gas \cite{Bogoliubov}  and to
the one of the Beliaev  equations for Bose superfluids \cite{Mul92,Bel58}.
With respect to those theories the present approach makes use of a finite
ranged and momentum  dependent effective interaction. The same theory gives
a reliable  description of  surface modes (ripplons) both at small and high
momenta  \cite{Las95}. In the vacuum the equations of motion  coincide with
the  Schr\"odinger equation for the free atom wave function  $f(z)$, while 
$g(z)$ vanishes.  

The equations for $f(z)$ and $g(z)$ are solved numerically in a finite box. 
We work in a slab geometry, that is, with the liquid confined within two
parallel  surfaces, the slab being centered in the box. The box size is of
the order of  $100 \div 150$ \AA\ and the slab thickness is typically $50
\div 100$ \AA\  to simulate sufficiently well  the semi-infinite medium. The
space along  $z$ is represented by a mesh of points, with step $0.1 \div
0.15$ \AA.  The  Hartree equation (\ref{eq:hartree}) for the  ground state
is solved  by   diffusing a trial wave function in  imaginary time.  After
several iterations  the procedure converges to the density profile which
minimise the energy  functional. The minimisation takes $2 \div 5$ hours of
CPU-time on a  workstation, depending on the accuracy required.  The same
calculation  provides the self-consistent mean field entering the static
Hamiltonian (\ref{eq:h0}). Then we expand the linearised time dependent wave 
function $\delta \Psi$  on the basis of eigenstates of  $H_0$.   By
this  way the  equations of motion for $f$ and $g$ become a matrix equation
for  the coefficients of the expansion. We take a basis of  $100$ states or
more  to  cover all the relevant part of the spectrum, so that the matrix to 
diagonalise is at least of dimension $100 \times 100$. The solution of the 
equations of motion for a given value of parallel wave vector $q_x$ takes 
about two hours of CPU-time on a workstation.  The output is a  set of 
discrete eigenenergies and the corresponding functions $f(z)$ and $g(z)$. 
By solving for different values of $q_x$ one can span the 
whole spectrum. An example is shown in Fig.~\ref{fig:spectrum2}, where 
we plot the energies of the excitations of a slab $80$\AA\ thick 
in a box of $150$\AA.   One recognises clearly the main features already
reported in  Fig.~\ref{fig:spectrum}, namely the phonon-roton dispersion and 
threshold for free atom states. The spacing in energy depends on the 
thickness of the slab and of the free space between the slab surface and the 
confining box. In the limit of infinite slab thickness the excitations in the 
liquid will reduce to a continuum of states, as already discussed for 
Fig.~\ref{fig:spectrum}.  In Fig.~\ref{fig:spectrum2} one also notices lower 
branches below the roton energy $\Delta$ and the phonon dispersion. 
They correspond to surface modes and a discussion about them has been 
already reported in Ref.~\cite{Las95}. The function $f$ and $g$,  at 
given $q_x$ and $\omega$, will take contributions from the elementary 
modes propagating inside and outside the liquid at different angles with 
respect to the surface.  The contribution of each 
elementary mode (phonon, $R^+$ and $R^-$ roton, atom, surface mode) is 
characterised by the appropriate dispersion law and can be identified by 
looking at the form of  $f(z)$ and $g(z)$ and their Fourier transforms. 
An example is given in Fig.~\ref{fig:exampleofstate}, where we show the
form of $f(z)$ and $g(z)$ for one of the states of the spectrum in
Fig.~\ref{fig:spectrum2}. The amplitude and the wave vector associated
with the free atom are easily extracted from the form of $f(z)$
close to the  boundary. The amplitudes and wave vectors of rotons can be 
calculated by looking at the Fourier transforms of $f(z)$ and $g(z)$ in the
liquid. In the lower part of Fig.~\ref{fig:exampleofstate} we show the
Fourier transform of $f(z)$. In the limit of infinite slab thickness, the 
function $f(q_z)$ would reduce to the sum of two $\delta$-functions of the
form $f_\pm \delta(q_z -q_{z\pm})$, corresponding to $R^\pm$ rotons. Since
the Fourier  analysis is restricted to a finite box within the slab,
$f(q_z)$ is actually the convolution of those $\delta$-functions with the
transform of a step function; this convolution produces the
oscillations visible in the figure. The amplitudes $f_\pm$ and the wave
vectors $q_{z\pm}$ are extracted with great accuracy by fitting $f(q_z)$. 
In the case of Fig.~\ref{fig:exampleofstate}b the best fit is  
indistinguishable from $f(q_z)$. 

All these numerical results are used as input in the calculation of the flux 
of excitations in the scattering processes at the surface, as explained  in
the  following section. 

\section{Flux of elementary excitations}
\label{sec:3} 

In the standard theory of scattering the probability associated with a given 
output channel is written as the ratio between the outgoing flux in that
channel  and the  incoming flux.  In the case of quantum evaporation 
the free  surface acts as scattering region and one has to consider 
the flux of  elementary excitations, i.e., rotons, phonons and atoms.  The
solutions of  the equations of motion provide the fluctuations of the
density and current of helium atoms  expressed by means of
functions $f(z)$ and $g(z)$ entering $\delta \Psi$.   These functions are
different inside and outside the liquid and  the boundary  produces the
appropriate matching between the excitations of the bulk  liquid and the free
atoms states in vacuum.  Now one has to relate the  asymptotic form of these
functions  with the flux of incoming and  outgoing excitations in a given
scattering process. We proceed in two steps.  First we introduce the concept
of current of elementary excitations within the  formalism of
linearised time dependent mean field theory.  Then we  represent the
scattering processes by taking linear combinations of  stationary solutions
of the TDDF equations. 

Far away from the surface and  for a given energy $\hbar \omega$ the 
function $\delta \Psi$ is the sum of plane waves of the form 
\begin{equation}
\delta \Psi ({\bf r}) 
= \sum_j \left[ f_j({\bf r}) e^{-i \omega t}
+ g^*_j({\bf r}) e^{i \omega t} \right] 
= \sum_j \left[ f_j  e^{i ( {\bf q} \cdot {\bf r} - 
\omega  t)} + g_j e^{- i({\bf q} \cdot {\bf r} - \omega t)} \right]
\label{eq:planewaves}
\end{equation}
where the index $j=a,+,-$ refers to the type of excitations (atoms, $R^+$ 
and $R^-$ respectively) which contributes to $\delta \Psi$. Each  type of
excitation obeys its own dispersion law $\omega({\bf q})$ and contributes 
to $\delta \Psi$ with amplitudes $f_j$ and $g_j$. 

Well outside the liquid  $\delta \Psi$ describes free atoms. The
hole-particle part $g_a$  vanishes, while the atom density is given by
$\rho_a=f_a^2$. The density of current is given by 
\begin{equation}
{\bf j}_a = {\hbar {\bf q} \over m } f_a^2 = {\bf v}_a f_a^2 \ \ \  ,
\label{eq:ja}
\end{equation}
where ${\bf v}$ is the group velocity which, for  the free atoms, coincides
with  $\hbar {\bf q}/m $.  These are the usual definitions of  density
and current density for free particles.

Inside the liquid the same  definitions can be generalised to describe the 
elementary modes of the correlated system. The RPA (or, equivalently,
Bogoliubov) formalism provides a natural generalisation for the density of
elementary excitations which, in term of the components $f$ and $g$ of the
time dependent solution (\ref{eq:deltapsi1}), is given by
\begin{equation}
\rho^{ex}_j \equiv |f_j({\bf r})|^2 - |g_j({\bf r})|^2 = 
f_j^2 - g_j^2 \; .
\label{eq:rhoj}
\end{equation}
This expression is consistent with the well known ortho-normalisation
property of the RPA solutions:
\begin{equation}
\int \! d{\bf r} \ [ f_i^*({\bf r}) f_j({\bf r}) - 
 g_i^*({\bf r}) g_j({\bf r}) ] = \delta_{ij}  \; . 
\label{eq:orthonorm}
\end{equation}
The total density of elementary excitations, $\rho^{ex}=\sum_j \rho^{ex}_j$,
satisfies the equation of continuity:
\begin{equation}
{ \partial \rho^{ex} \over \partial t} +  \nabla \cdot {\bf j}^{ex} = 0  
\ \ \ . 
\label{eq:continuity}
\end{equation}
The quantity ${\bf j}^{ex}=\sum_j {\bf j}_j^{ex}$ is identified with the
current density of excitations, for which one expects the  result
\begin{equation}
{\bf j}^{ex}_j = {\bf v}_j \rho^{ex}_j \equiv {\bf v}_j (f_j^2 - g_j^2) 
\label{eq:jj}
\end{equation}
where ${\bf v}_j= \nabla_{\bf q} \omega(q)$ is the group velocity of the
$j$-th elementary excitation. The group velocity should not be  confused 
with the quantity $\hbar {\bf q}/m$ (which is the correct expression only
for free atoms) and actually becomes opposite to ${\bf q}$ in the $R^-$
range of wave vectors. In  Appendix B we show that the form
(\ref{eq:jj}) for the current density is fully consistent with the equations
for $\rho^{ex}_j$  derived starting from the solutions of
the TDDF equations. We note also that, when  $\rho^{ex}$ and ${\bf j}^{ex}$
are calculated by using the stationary solutions of the equations of motion,
the time derivative vanishes and the equation of continuity implies 
simply that the $z$-component of the current of elementary excitations 
is  equal inside and outside the liquid. This result is also connected
with the unitarity of the scattering matrix, as we will discuss in the next
section.

Equations (\ref{eq:rhoj},\ref{eq:continuity},\ref{eq:jj})  emphasise a
remarkable feature of the TDDF solutions.  In a dilute system they  reduce
to the usual expressions holding for free particles,
while in the interacting liquid they   include  the correlations associated
with  the propagation of the collective modes.   For instance in the long
wavelength phonon regime  ($q \to 0$)  the group  velocity coincides with
the sound velocity and $g \simeq f$.  

From the knowledge of the current of elementary excitations associated 
with each asymptotic solution  one can calculate 
the flux of incoming and outgoing excitations in a scattering process.  
Actually the output of our TDDF equations, as discussed in Sec.~\ref{sec:2}, 
are stationary solutions,  namely real functions $f(z)$ and $g(z)$ which 
vanish at the border of the computational box. These functions can be 
viewed as a combination of $R^+$ and $R^-$ rotons in the liquid and atoms 
outside,  travelling to and from the surface as shown schematically in the
upper part of Fig.~\ref{fig:schemeofscatt}.  Using Eq.~(\ref{eq:jj}) one
can  calculate the current associated with each component.  In order to
select a  single scattering process one can combine two or more stationary
solutions  at the same energy and parallel wave vector \cite{note3}. If
$f^{(1)} (z)$,  $f^{(2)} (z)$ and $f^{(3)} (z)$ are three  solutions of the
linearised TDDF equation at the same energy and parallel wave vector, then
the combination 
\begin{equation} 
f (z) =  f^{(1)} (z) + a_2 f^{(2)} (z) + a_3 f^{(3)} (z)
\label{eq:comb}
\end{equation}
is also a solution (the same holds for $g(z)$).  One can choose the
coefficients of  the combination in such a way that part of the incoming 
flux  vanishes. For  instance, if we want to represent one $R^+$ roton
impinging onto the  surface and reflecting as $R^+$ or $R^-$, or evaporating
an atom, we have  to combine the TDDF solutions in order to make the incoming
fluxes of  atoms and $R^-$ rotons vanish.  This provides two equations for
the complex  coefficients $a_2$ and $a_3$ in terms of the amplitudes
$f^{(\alpha)}$ and  $g^{(\alpha)}$ (the third coefficient  $a_1$  fixes the
normalisation of the wave function and its value is not relevant when one
calculates the   ratio of fluxes).  Once the coefficients are calculated,
one is left  with a  solution which represents a flux (current density
projected along  $z$) of  incoming $R^+$ rotons and a flux of evaporated
atoms and  reflected $R^+$  and $R^-$ rotons, as in the lower part of 
Fig.~\ref{fig:schemeofscatt}.   By definition, the evaporation  probability
is  given by   
\begin{equation}
P_{+a} =  { \text{ flux of evaporated atoms} \over
\text{flux of incoming}\ R^+ }  \; , 
\end{equation}
the normal reflection probability is 
\begin{equation}
P_{++} = { \text{ flux of reflected}\ R^+ \over
\text{flux of incoming}\ R^+ } 
\end{equation}
and the mode-change reflection is 
\begin{equation}
P_{+-} =  { \text{ flux of of reflected}\ R^- \over
\text{flux of incoming}\ R^+ }  \; .
\end{equation}
The explicit formulae 
for this scattering process, including the geometry of the slab and the box, 
are given in Appendix C.  By combining the stationary solutions with 
different coefficients  one can select the process in which the incident 
excitation is a $R^-$ roton. In this case the ratios of the outgoing fluxes
and the incoming $R^-$ flux  provide the probabilities $P_{-a}$
(evaporation),  $P_{--}$ (normal reflection) and  $P_{-+}$ (mode-change
reflection). Similarly, by selecting the process with an incident atom one
gets the probabilities $P_{aa}$ (reflection), $P_{a+}$ (condensation in
$R^+$) and  $P_{a-}$ (condensation in $R^-$).  In region III of
the spectrum in Fig.~\ref{fig:spectrum} these are the only  one-to-one
processes available. Generalisation to the other regions of  the
spectrum is straightforward.

The method we have used in Ref.~\cite{Dal95} to derive the evaporation 
and reflection probabilities  is simpler but less general. It corresponds to
the  case when one of the state in the  linear combination has vanishing 
amplitude outside the liquid, due to destructive interference.  
The existence of such  {\sl resonant states} simplifies the calculation of
the  probabilities $P_{ij}$.   However, the more general method used in the
present work is not  restricted to special states and provides results on
a wider range of energy. Furthermore it can be easily extended to  describe
more scattering channels.

\section{The Scattering Matrix}
\label{sec:4}

The scattering matrix $S$ characterises a scattering process through the 
definition  $\Psi_{out}=S \Psi_{in}$. The unitarity and the time reversal 
symmetry ($t \to -t$) of $S$ imply 
\begin{equation}
S^\dagger = S^{-1} 
\label{eq:sdagger}
\end{equation} 
and 
\begin{equation}
S^*= S^{-1} \; ,
\end{equation}
respectively.  Combining these two conditions one finds that the scattering
matrix elements must satisfy the general property 
\begin{equation}
S_{ij}=S_{ji} \; . 
\label{eq:symmetryij}
\end{equation}
The scattering matrix elements are related to the probabilities 
$P_{ij}$ by
\begin{equation}
P_{ij}=|S_{ij}|^2 \; .
\label{eq:pijsij}
\end{equation}
Equation (\ref{eq:symmetryij}) then implies that the evaporation,
reflection and condensation rates, introduced in the previous section, 
reduce to six quantities to be calculated: 
$P_{+a}$, $P_{-a}$, $P_{++}$, $P_{--}$, $P_{+-}$ and $P_{aa}$. 
Furthermore the unitarity condition takes also the
form $S^*S=1$, or 
\begin{equation}
|S_{+a}|^2 + |S_{++}|^2 + |S_{+-}|^2 = 1  
\label{eq:unitarity1}
\end{equation}
\begin{equation}
|S_{-a}|^2 + |S_{--}|^2 + |S_{-+}|^2 = 1 
\label{eq:unitarity2}
\end{equation}
\begin{equation}
|S_{+a}|^2 + |S_{-a}|^2 + |S_{aa}|^2 = 1 \; . 
\label{eq:unitarity3}
\end{equation}
These conditions express the conservation of flux in the three processes 
where  the incoming  excitation is a $R^+$ roton, a $R^-$ roton and an 
atom, respectively. They reflect the fact that the present approach does
not include inelastic channel. The unitarity of the $S$ matrix gives three
additional constraints: 
\begin{equation}
S^*_{aa} S_{+a} + S^*_{a+} S_{++} + S^*_{a-} S_{+-}    = 0 
\label{eq:unitarity*1}
\end{equation}
\begin{equation}
S^*_{aa} S_{-a} + S^*_{a+} S_{-+} + S^*_{a-} S_{--}    = 0 
\label{eq:unitarity*2}
\end{equation}
\begin{equation}
S^*_{+a} S_{-a} + S^*_{++} S_{-+}  + S^*_{+-} S_{--}   = 0 \; . 
\label{eq:unitarity*}
\end{equation}
All these conditions for the scattering matrix elements can be used to
obtain useful  checks for the results of the numerical calculation.  

\section{Results}
\label{sec:5} 

The main results of the numerical analysis are shown in 
Fig.~\ref{fig:results}, where we  report the probabilities
$P_{ij}=|S_{ij}|^2$ as a  function of  energy.  The data
are obtained with several values of the  parallel wave vector $q_x$ in order
to span the phonon forbidden region (region III of the  spectrum
in Fig.~\ref{fig:spectrum}). The data at lowest energies in 
Fig.~\ref{fig:results} correspond to $q_x=0.6$ \AA$^{-1}$, the ones at
highest energy to $q_x=0.85$ \AA$^{-1}$. From the knowledge of the orthogonal
wave vector $q_z$ one extracts also the incidence angle. For $R^+$ and $R^-$ 
rotons it is approximately $15^\circ \div 20^\circ$ and  $20^\circ \div
30^\circ$, respectively, while for atoms is $50^\circ \div 90^\circ$. The
range of incidence angles for rotons is  rather narrow, so that a discussion
about the angular dependence of the evaporation and reflection probabilities
is not significant in region III of the spectrum. Further work is planned to
explore the angular dependence including region II.  The error bars arise
from fluctuations of the numerical results. They originate mainly from the
fact that the states in the linear combination (\ref{eq:comb}) may be not
enough linearly independent. As explained at the end of Appendix
\ref{app:c}, this implies  an  amplification of
small numerical uncertainties in the values of amplitudes, wave vectors and
other parameters entering the expressions for $P_{ij}$. 

Looking at Fig.~\ref{fig:results} one notices that the unitarity conditions 
(\ref{eq:unitarity1}-\ref{eq:unitarity3})  are 
satisfied within the error bars. Actually each point with error bar
corresponds to several combinations of states at the same energy.  The
unitarity condition can be checked for each combination separately. It turns
out that it is satisfied within about $5$\% when the states in the
combination are sufficiently linearly independent. An example of numerical
values is  given in Table \ref{tab:combinations} for states at energy
$10.87$ K. The results of different combinations of four states are given.
The four states correspond to symmetric and antisymmetric solutions for two
slabs ($62$ and $63.7$ \AA\ thick). One notes that the values of the
probabilities $P_{ij}$ extracted form different combinations are
close each other. The unitarity conditions, which correspond to $\Sigma=1$
in the notation of the table, is satisfied within $5$\% in all cases.
Furthermore one has $P_{ij} \simeq P_{ji}$ as expected from
Eq.~(\ref{eq:symmetryij}). 

The evaporation probabilities for $R^+$ and $R^-$ rotons behave quite
differently.  The one for  $R^+$ rotons grows from $0$ to $1$ by increasing
the energy. A similar behavior has been recently observed in experiments at
normal incidence in the same range  of energy\cite{For96}. Conversely  the
$R^-$ rotons are less effective in evaporating atoms. The ratio between the
two probabilities has to be $1$ at the energy $\Delta$ of the roton minimum
where the two excitations become identical. As soon as the energy grows the
ratio $P_{+a}/P_{-a}$ increases and becomes large. Results for
$P_{+a}/P_{-a}$  are shown in  Fig.~\ref{fig:p+/p-}.

At low energies, approaching $\Delta$, the mode-change is dominant in the
reflection processes for rotons. This is also consistent with symmetry
arguments (see discussions in Ref.~\cite{Las95}). At higher energies the
mode-change becomes less probable.  An incident $R^-$ rotons has low
probability of evaporating atoms, as discussed above, and it reflects as
$R^+$ at low energy and as $R^-$ at high energy. The behavior of an incident
$R^+$ roton at high energy is completely different: the normal reflection as
$R^+$ is found to be practically zero everywhere. $R^+$ rotons  prefer to
evaporate atoms or to reflect as $R^-$, the sum of the two probabilities
being almost $1$. This fact can be explained qualitatively by considering
the change of momentum in a normal mode reflection. This should be twice
$\hbar q_z$, where $q_z$  in this part of the spectrum is of the order of $2$
\AA$^{-1}$. But it is difficult to imagine a mechanism for  absorbing  a
momentum of the order of $4$ \AA$^{-1}$  at the surface, the latter being
quite smooth  (the Fourier transform of the density profile has vanishing
components above $2$ \AA$^{-1}$). On the contrary, the momentum transfer for
reflections including $R^-$ rotons is at least a factor two smaller and it is
expected to be more probable. 

The fact that $P_{++}$ is almost zero everywhere has an interesting
consequence. If we assume $S_{++}=0$ and use the unitarity conditions
(\ref{eq:unitarity1}-\ref{eq:unitarity*}), we can express all the
probabilities through a single parameter, for instance  $P_{+a}$. Simple
algebra gives (see Appendix D): 
\begin{equation}
P_{+-} = 1 - P_{+a}
\label{eq:special1}
\end{equation}
\begin{equation}
P_{--} = P^2_{+a}
\label{eq:special2}
\end{equation}
\begin{equation}
P_{aa} = P_{+-}^2 = (1 - P_{+a})^2
\label{eq:special3}
\end{equation}
\begin{equation}
P_{-a} = P_{-+} P_{+a} = (1 - P_{+a})  P_{+a}
\label{eq:special4}
\end{equation}
Note that the maximum value of $P_{-a}$, from the last equation, is 
obtained for $P_{+a}=1/2$ and thus the following inequality holds:
\begin{equation}
 P_{-a}   \le {1 \over 4} \; .
\end{equation}
All the above relations, which follow from the assumption  $S_{++}=0$,
are  well satisfied by the numerical results in Fig.~\ref{fig:results}. 

Finally, the results for the reflection probability for incident atoms are
more intriguing. One knows from the experiments that the reflection
probability is less than $1$\% apart from very small values of $q_z$ where
the long range Van der Waals interaction dominates.   Our results for the
reflection probability are consistent with the experimental results only at
high energy, where  $P_{aa}$ drops to zero. At lower energy the roton
channels become less and less active in the condensation process and the
atom is reflected with large probability.  This behavior is  mainly the
consequence of the fact that in the range of angles considered in the
present work phonons cannot be excited. As soon as phonons are allowed
(region I and II in Fig.~\ref{fig:spectrum}) they are expected to provide
the natural channel for the condensation of the incoming atoms. It has been
also suggested that, as soon as the atom has enough energy to feel the
external part of the density profile, it dissipates its energy by producing
many low energy ripplons. This mechanism is not accounted for in our
one-to-one theory.

\section{Conclusions and perspectives} 
\label{sec:conclusion}

In this paper we have investigated the evaporation and condensation
processes taking place at the free surface of superfluid helium. 
The main achievements of the present work concern both the  formal
development of scattering theory, in the framework of linearised time 
dependent density functional formalism, and the numerical analysis,  which 
provides first systematic results for the evaporation,
condensation and reflection rates. The most relevant steps and achievements
of our approach are here summarised:

\begin{itemize}

\item i) The equations of motion of linearised time dependent density
functional theory have been explicitly solved in the presence of the free
surface using the Orsay-Trento phenomenological density functional. 
By taking linear combinations of different solutions carrying the same
energy and momentum parallel to the surface, we have selected the
relevant scattering processes (incoming atom reflected into an atom or
condensed into a $R^+$ or $R^-$ roton, etc.).

\item ii) We have identified, in the framework of the time dependent density 
functional scheme, the current of elementary excitations which
obeys the equation of continuity (\ref{eq:continuity}) and allows one to
evaluate the rates of evaporation and reflection. This current is determined
by the group velocity and by the density of elementary excitations
for which the familiar RPA expression (\ref{eq:rhoj}) holds. The resulting
formalism has been shown to be consistent with the general properties of the 
scattering matrix, following from unitarity and time reversal invariance.

\item iii) The numerical results, restricted in the present work to
the region where only condensation of atoms into rotons is possible
(region III of Fig.~\ref{fig:spectrum}), reveal that rotons $R^+$ are
systematically more effective than rotons $R^-$  in the evaporation
mechanism. An interesting feature emerging from our analysis, is that the 
$P_{++}$ scattering probability is extremely small at all energies. This
behavior, with the help of the formal properties of the scattering matrix,
allows one to  express all the evaporation and reflection rates in terms
of a single parameter [see Eqs.~(\ref{eq:special1}-\ref{eq:special4})]. 

\item iv) A major discrepancy between our predictions and the experimental
data concerns the atomic reflectivity which is predicted to be significantly
large when the energy approaches the roton minimum. 
This behavior is easily understood since our theory accounts only for
one-to-one processes where condensation into  ripplons is excluded
by the conservation laws. Due to the absence of phonon excitations, which
are  excluded in region III, no active excitations are allowed in the
condensation process when the energy approaches the roton minimum with the 
consequent full reflection of the incident atom.  

\end{itemize}

Natural developments of the present work  will include:

\begin{itemize}

\item The study of the scattering process in different ranges of
energy and incident angles, where phonons can participate in the 
scattering process. 

\item The study of multi-ripplon excitations whose effects 
can be taken into account through the use
of an optical-type potential in the time dependent equations.

\end{itemize}

Work in these directions is in progress.

\acknowledgements

We are indebted to A.F.G. Wyatt, A.C. Forbes, D.O. Edwards and M. Guilleumas
for many useful discussions.  L.P.  thanks the hospitality of the
Dipartimento di Fisica at the University of Trento  as well as of the
Institute for Condensed Matter Theory at the University of Karlsruhe and
support of the Alexander von Humboldt Foundation.

\appendix

\section{Orsay-Trento density functional}
\label{app:a}

We use the  phenomenological density  functional of Ref.~\cite{Dal95b}. It 
has the form: 
\begin{equation}
E \ = \ E^{(kin)} [\rho,{\bf v}] +  E^{(c)}[\rho] +  
E^{(bf)}[\rho,{\bf v}] \; ,
\label{app:ot}
\end{equation}
where $\rho$ and ${\bf v}$ are the density and velocity of the atoms, 
respectively. The first term is the kinetic energy of the non interacting
bosons, \begin{eqnarray}
E^{(kin)} [\rho,{\bf v}] 
        \ &=& \ \int \! d{\bf r} \ {\hbar^2 \over 2m}|\nabla \Psi({\bf r})|^2 
\nonumber  \\
            &=& \int\! d{\bf r} \bigg\{ 
                {\hbar^2 \over 2m} (\nabla \sqrt{\rho})^2 
		+ {m \over 2} \rho({\bf r}) |{\bf v}({\bf r})|^2 \bigg\} \;  ;
\label{app:ekin}
\end{eqnarray}
the correlation energy $E^{(c)}$ is given by
\begin{eqnarray} 
E^{(c)}[\rho]  &=& \ \int \! d{\bf r} \Big\{
        {1\over 2} \int \!  d{\bf r}' \ \rho({\bf r}) V_l(|{\bf r}- {\bf
r}'|) \rho({\bf r}') 
	\ + \ {c_2 \over 2} \rho({\bf r})  (\bar \rho_{{\bf r}})^2 
        \ + \ {c_3  \over 3} \rho({\bf r})  (\bar \rho_{{\bf r}})^3 
\nonumber
\\     &-& {\hbar^2 \over 4m} \alpha_s \int \! d{\bf r}' \ F(| {\bf r} 
-{\bf r}' |)  
	       \left(1-{\rho({\bf r}) \over \rho_{0s}} \right) 
	       \nabla \rho({\bf r}) \cdot \nabla \rho({\bf r}')
	 \left(1-{ \rho({\bf r}') \over \rho_{0s}} \right) \Big\}  \;   ; 
\label{app:ec}
\end{eqnarray}
finally,  the backflow energy $E^{(bf)}$ is
\begin{equation}  
E^{(bf)} [\rho,{\bf v}] \  = \  
 - {m \over 4} \int\! \int \! d{\bf r} d{\bf r}' \ V_J (|{\bf r}-{\bf r}'|) \ 
\rho({\bf r}) 
 \rho({\bf r}') \  \left[ {\bf v}({\bf r}) -{\bf v}({\bf r}') \right]^2 \; . 
\label{app:ebf}
\end{equation}
The first term in the kinetic energy, which depends
on gradient of the density,   is a quantum pressure;  it  
corresponds to  the zero temperature kinetic energy of  
non-interacting bosons of mass $m$. The two-body interaction $V_l$ in 
the correlation energy $E^{(c)}$ is the Lennard-Jones 
interatomic potential, with the standard parameters  $\alpha 
=2.556$ \AA \ and $\varepsilon=10.22$ K, screened at short 
distance ($V\equiv 0$ for $r<h$, with $h=2.1903$\AA). The two  
terms with the parameters  $c_2 = -2.411857 \times 10^4$ K \AA$^6$ 
and $c_3=1.858496 \times 10^6$ K \AA$^9$ 
account phenomenologically for short range correlations  
between atoms. The  weighted density $\bar \rho$ is the 
average of $\rho ({\bf r})$ over a sphere of radius $h$. Those 
terms are very similar to the functional of Ref. \cite{Dup90}.
The last term in $E^{(c)}$,  depending on the gradient of the 
density in different points,  has been added in 
order to improve the description of the static response function
in the roton region. The function $F$ is a 
simple Gaussian, $F (r)= \pi^{-3/2} \ell^{-3} \exp(-r^2/\ell^2)$
with  $\ell=1$~\AA, while  $\alpha_s=54.31$~\AA$^3$ and   $\rho_{0s}
=0.04$ \AA$^{-3}$. The energy $E^{(bf)}$ contains an effective 
current-current interaction accounting for backflow-like correlations.
In Ref.~\cite{Dal95b} the simple parametrisation
\begin{equation}
 V_J (r) = (\gamma_{11} + \gamma_{12} r^2) \exp(-\alpha_1 r^2) 
         + (\gamma_{21} + \gamma_{22} r^2) \exp(-\alpha_2 r^2) \ \ \ ,
\label{app:vj}
\end{equation}
was chosen in order to reproduce the phonon-roton dispersion in 
bulk liquid. The parameters are given in Table \ref{table}.

\section{Equation of continuity and current of elementary excitations} 
\label{app:b}

In this appendix we show that definition (\ref{eq:jj}) for the current
of elementary excitations is consistent with equation (\ref{eq:leastaction})
holding in time dependent density functional theory. In order to 
make the demonstration more transparent  we use a simplified version of the 
energy functional of the form  
\begin{equation}
 {\cal H} = {\hbar^2 \over 2 m}  |\nabla \Psi({\bf r}_1)|^2 +
\int\! d{\bf r}_2 V(|{\bf r}_1-{\bf r}_2|) \rho({\bf r}_1)\rho({\bf r}_2) \; , 
\label{app:functional}
\end{equation}
where $V$ is a generic two-body interaction.  The generalisation to the 
more complex Orsay-Trento functional is conceptually straightforward but 
rather tedious in practice. 

Taking variations with respect $\Psi^*$ in  Eq.~(\ref{eq:leastaction}) 
one gets
\begin{equation}
\left[ 
-i \hbar {\partial \over \partial t} - {\hbar^2 \over 2m}  \nabla^2  +
\int \! d{\bf r}_2  V(|{\bf r}_1-{\bf r}_2|)  \rho({\bf r}_2)  - \mu  
\right] \Psi({\bf r}_1,t) = 0  \; .
\label{app:hpsi}
\end{equation}
Here we are interested in the solutions well inside the liquid, where the 
ground state density $\rho_0$  is a constant and the above equation  yields
\begin{equation}
\mu = \rho_0 \int\! dr  V(r) \; . 
\label{app:mu}
\end{equation}
As in Eq.~(\ref{eq:deltapsi1}) we take a linear expansion of $\Psi$ to 
describe  the excited state. Let us include the time dependence  in the
definitions of $f$  and $g$ in  the form 
\begin{equation}
f({\bf r},t) = f  e^{i({\bf q} \cdot {\bf r} - \omega t)} \ \ \     ;  \ \ \
g({\bf r},t) = g e^{i({\bf q} \cdot {\bf r} - \omega t)} \;  ,
\label{app:planew}
\end{equation}
so that 
\begin{equation}
\delta \Psi ({\bf r},t) = f({\bf r},t) + g^*({\bf r},t) \; . 
\label{app:deltapsi}
\end{equation}
Inserting this expression in Eq.~(\ref{app:hpsi}) and isolating the
terms oscillating with  positive and  negative frequencies, one
gets  two coupled equations for $f$ and $g$: 
\begin{equation}
- i \hbar {\partial f(1) \over \partial t} -  
{\hbar^2 \over 2m}  \nabla^2 f(1) 
+ \rho_0 \int\! d{\bf r}_2 V(r_{12})   [f(2) + g(2)]
= 0  
\label{app:couple1}
\end{equation}
\begin{equation}
i \hbar {\partial g(1) \over \partial t} -  
{\hbar^2 \over 2m}  \nabla^2 g(1) 
+ \rho_0 \int\! d{\bf r}_2 V(r_{12})  [f(2) + g(2)]
= 0  
\label{app:couple2}
\end{equation}
where we have used the short notation $(1)\equiv ({\bf r}_1,t)$ and 
$(2)\equiv ({\bf r}_2,t)$.  Now, take $f(1)$ times the complex conjugate of  
Eq.~(\ref{app:couple1})  and subtract $f^*(1)$ times Eq.~(\ref{app:couple1}) 
itself. Similarly take $g(1)$ times the complex conjugate of 
Eq.~(\ref{app:couple2})  and substract $g^*(1)$ times
Eq.~(\ref{app:couple2}).  Summing up all the terms, one gets
\begin{eqnarray}
i \hbar {\partial \over \partial t} ( |f(1)|^2 - |g(1)|^2)
\ &=& \ {\hbar^2 \over 2m}  \bigg[  f(1) \nabla^2 f^*(1) 
- f^*(1) \nabla^2 f(1) + g(1) \nabla^2 g^*(1) - g^*(1) \nabla^2 g(1) \bigg]
\nonumber  \\
 &-&\rho_0 \int\! d{\bf r}_2  V(r_{12}) \bigg\{ f(1)f^*(2) - f^*(1)f(2) 
+ f(1)g^*(2) - g^*(1)f(2)
\nonumber \\
&+& g(1)f^*(2)- f^*(1)g(2) + g(1)g^*(2) - g^*(1)g(2) \bigg\} \; . 
\label{app:big}
\end{eqnarray}
This equation has the form of an equation of continuity if the quantity
\begin{equation}
\rho^{ex}= |f(1)|^2-|g(1)|^2 = f^2-g^2
\label{app:rhoex}
\end{equation} 
is interpreted as the  density  of elementary excitations  and if the
right hand side is identified with  $-i\hbar \nabla \cdot  {\bf j}^{ex}$. 
Now we prove that the quantity ${\bf j}^{ex}$ has indeed the form ${\bf
j}^{ex} = {\bf v}  \rho^{ex}$, where ${\bf v}$ is the group
velocity of the excitation.  To this purpose  we rewrite properly  the
various terms in Eq.~(\ref{app:big}).  We  begin with the one coming from
the kinetic energy.  All  terms with the  laplacian can be written in the
form  $-i \hbar \nabla \cdot [{\bf j}^{ex}]_{kin}$,  with  
\begin{equation} 
[{\bf j}^{ex}]_{kin} = - {\hbar \over 2mi} \left[ f(1)\nabla
f^*(1) - f^*(1) \nabla f(1) + g(1) \nabla g^*(1) - g^*(1) \nabla g(1)
\right]  \; .  
\label{app:jkin}
\end{equation}
We remember that  $f(1)$ and $g(1)$ are plane waves, as in
Eq.~(\ref{app:planew}).  Thus  the expression for  $[{\bf j}^{ex}]_{kin}$,
in terms of the real amplitudes $f$ and $g$, becomes 
\begin{equation}
[{\bf j}^{ex}]_{kin} = { \hbar {\bf q} \over m} ( f^2+ g^2)  \; , 
\label{app:jkin2}
\end{equation}
To manage the terms containing the interaction is less trivial. Let us use 
the Fourier decomposition 
\begin{equation}
f({\bf r},t) = \int\! d{\bf q}_1 f({\bf q}_1,t) e^{i {\bf q}_1 \cdot {\bf r}}
\label{app:fourier} 
\end{equation}
and take the limit $f({\bf q}_1) = f  \delta ({\bf q}_1 - {\bf q})$ at the 
end.   We can rewrite the first two terms with the interaction in 
Eq.~(\ref{app:big}) as 
\begin{eqnarray}
\int\! d{\bf r}_2  V(r_{12}) \bigg\{ f(1)f^*(2) - f^*(1)f(2) \bigg\} 
& = &
\int\! \int \! d{\bf q}_1 d{\bf q}_2 \ 
e^{  i  ( {\bf q}_1-{\bf q}_2 ) \cdot {\bf r}_1 }  
\nonumber  \\ 
& \times &  f({\bf q}_1,t) f^*({\bf q}_2,t) ( V({\bf q}_2) - V ({\bf q}_1))
\label{app:firsttwo}
\end{eqnarray}
where $V({\bf q})$ is the Fourier transform of the interaction. By doing the 
same with the terms containing  $g$, one gets 
\begin{eqnarray}
- i \hbar \nabla \cdot [{\bf j}^{ex}]_{pot} 
& = & - \rho_0 \int\! \int \! d{\bf q}_1 {\bf q}_2 
 e^{i({\bf q}_1-{\bf q}_2)\cdot {\bf r}_1} ( V({\bf q}_2) - V ({\bf q}_1)) 
\bigg[ f({\bf q}_1,t) f^*({\bf q}_2,t) \nonumber \\
& + &   
g({\bf q}_1,t) g^*({\bf q}_2,t) + f({\bf q}_1,t) g^*({\bf q}_2,t)
 + g({\bf q}_1,t) f^*({\bf q}_2,t) \bigg]  \; .
\label{app:allterms}
\end{eqnarray}
Now, by considering $f({\bf q})$ and $g({\bf q})$ as delta functions, 
one gets
\begin{equation}
[{\bf j}^{ex}]_{pot} = \rho_0 (f+g)(f^*+g^*) \nabla_q V(q)  \; . 
\label{app:jexpot}
\end{equation}
and summing $[{\bf j}^{ex}]_{pot} $ to $[{\bf j}^{ex}]_{kin}$ one
finally obtains the result 
\begin{equation}
{\bf j}^{ex}= {\hbar {\bf q} \over m}  ( f^2+ g^2) + \rho_0 (f+g)(f^*+g^*)
\nabla_q V(q)   
\label{app:jex1}
\end{equation}
for the total current. 
At this point it is convenient to rewrite Eq.~(\ref{app:jex1}) in terms of
the  dispersion law $\omega(q)$ coming from the solutions of the equations
of  motion (\ref{app:couple1}-\ref{app:couple2}) which,  for 
plane waves  of the form (\ref{app:planew}) and after some algebra, yield
\begin{equation}
f+g = {\hbar q^2 \over 2 m \omega} (f-g) \; , 
\label{app:re1}
\end{equation}
\begin{equation}
f^2+g^2 =  {2m \over \hbar q^2 \omega} \left[ \omega^2 + \left({\hbar 
q^2 \over 2m}\right)^2 \right] (f^2 - g^2)  \;  , 
\label{app:res2}
\end{equation}
with 
\begin{equation}
\omega^2(q) =  { q^2 \over 2m} \left( {\hbar^2 q^2 \over 2m} + 2 \rho_0 
V(q) \right) \; . 
\label{app:dispersion}
\end{equation}
Inserting these results in Eq.~(\ref{app:jex1}) one finally finds
\begin{equation}
{\bf j}^{ex} = {\bf v} ( f^2 - g^2)  = {\bf v} \rho^{ex} \; , 
\label{app:jex}
\end{equation}
where ${\bf v}$ coincides with  the usual definition of the group
velocity: 
\begin{equation}
{\bf v} = \nabla_q \omega(q)  \; . 
\label{app:v}
\end{equation}
This is the desired definition of the current density, in terms of which 
Eq.~(\ref{app:big}) has the form of the equation of continuity
(\ref{eq:continuity}): 
\begin{equation} {\partial  \rho^{ex} \over \partial
t} +  \nabla \cdot  {\bf j}^{ex} = 0   \; . 
\label{app:continuity}
\end{equation}

\section{Determination of the  scattering rates} 
\label{app:c}

We give an example of derivation of the probabilities $P_{ij}$ 
in the case of $R^+$ rotons coming to the surface and producing 
evaporated atoms and reflected $R^+$ and $R^-$ rotons. The geometry we 
use is the one in Fig.~\ref{fig:geometry}: the surface is at $z=0$,  the
center  of the slab at $z=L$ and the left border of the box at $z=-b$. The
solutions of  our TDDF equations are real functions which vanish at $z=-b$.
They can  be either symmetric and antisymmetric with respect to the
center of the slab ($z=L$). If $b$ and $L$  are sufficiently
large, the functions $f(z)$ and $g(z)$, characterising the changes $\delta
\Psi$ as in Eq.~(\ref{eq:linearised}),  close to the boundary  and near the
origin are simple sinusoidal  functions. In particular, close to $z=-b$ one
has a free atom state of the form  \begin{equation} f(z) \simeq  f_a \sin
[q_a (z+b)]   = {f_a \over 2 i} e^{i q_a b} e^{i q_a z} -  {f_a \over 2 i}
e^{-i q_a b} e^{-i q_a z}   \label{app:sinout}
\end{equation} 
and $g=0$.  Close to $z=L$, inside the liquid, $f(z)$ is the sum 
of two oscillating functions of the form   
\begin{equation}
f(z) \simeq  f_\pm \sin [q_\pm (z-L)]   = {f_\pm \over 2i } e^{-i q_\pm L} 
e^{i q_\pm z} - {f_\pm \over 2i } e^{ i q_\pm L} e^{-i q_\pm z}  
\label{app:sinin}
\end{equation} 
in the antisymmetric case (a cosinus in the symmetric case) and the  
same for $g$. Here, for simplicity, the symbol $q$,    for
wave vectors, denotes the $z$-component only.  The Fourier analysis of the
solutions  of the TDDF equations provides the amplitudes  $f_a$ and $f_\pm$
for fixed  values of $L$ and $b$. Changing $L$ and $b$ one can find
different  solutions at the same energy and parallel wave vector.  Let us
choose two  symmetric states and an antisymmetric state and combine them in
the  form \begin{equation} 
f= f^{(1)} (z) + a_2 f^{(2)} (z) + a_3 f^{(3)} (z) \; ,  
\end{equation}
where the solution $f^{(\alpha)} $ corresponds to the parameters $L_\alpha$
and $b_\alpha$. The atomic part, well outside the slab, becomes
\begin{eqnarray} 
f_a  & = & {e^{i q_a z} \over 2i} \left[ f_a^{(1)}  e^{i q_a
b_1} +   a_2 f_a^{(2)}  e^{i q_a b_2} + a_3 f_a^{(3)}  e^{i q_a b_3} \right] 
\nonumber \\ & - &  {e^{-i q_a z} \over 2i} \left[f_a^{(1)}  e^{-i q_a b_1}
+   a_2 f_a^{(2)}  e^{-i q_a b_2} + a_3 f_a^{(3)}  e^{-i q_a b_3} \right] \;
,  \label{app:atom} \end{eqnarray}
the $R^+$ roton part, close to $z=L$, is 
\begin{eqnarray}
f_+  & = & {e^{i q_+ z} \over 2} \left[-i f_+^{(1)}  e^{i q_+ b_1} +  
a_2 f_+^{(2)}  e^{i q_+ b_2} + a_3 f_+^{(3)}  e^{i q_+ b_3} \right] 
\nonumber \\
& +&  {e^{-i q_+ z} \over 2} \left[ i f_+^{(1)}  e^{-i q_+ b_1} +  
a_2 f_+^{(2)}  e^{-i q_+ b_2} + a_3 f_+^{(3)}  e^{-i q_+ b_3} \right] 
\label{app:r+}
\end{eqnarray}
and the $R^-$ roton part is
\begin{eqnarray}
f_-  & = & {e^{i q_- z} \over 2} \left[-i f_-^{(1)}  e^{i q_- b_1} +  
a_2 f_-^{(2)}  e^{i q_- b_2} + a_3 f_-^{(3)}  e^{i q_- b_3} \right] 
\nonumber \\
& + &  {e^{-i q_- z} \over 2} \left[ i f_-^{(1)}  e^{-i q_- b_1} +  
a_2 f_-^{(2)}  e^{-i q_- b_2} + a_3 f_-^{(3)}  e^{-i q_- b_3} \right] \; .
\label{app:r-}
\end{eqnarray}
The  quantities in brackets represent the  amplitudes of   incoming and
outgoing excitations and  one can choose the coefficients  $a_2$ and $a_3$
in  order to make two of them vanish.  As said at the  beginning, we show
here the case of  an incident roton $R^+$. So, we must  have zero amplitude
for incoming atoms and $R^-$ rotons (see Fig.~\ref{fig:schemeofscatt}).
Keeping in  mind that $R^-$ rotons travel with negative group velocity, one
obtains  the following algebraic equations for $a_2$ and $a_3$: 
\begin{equation} f_a^{(1)}  e^{i q_a b_1} +  a_2 f_a^{(2)}  e^{i q_a b_2} +
a_3 f_a^{(3)}  e^{i q_a  b_3} = 0 \label{app:zero1}
\end{equation}
\begin{equation}
i f_-^{(1)}  e^{-i q_- b_1} +   
a_2 f_-^{(2)}  e^{-i q_- b_2} + a_3 f_-^{(3)}  e^{-i q_- b_3} = 0
\; . 
\label{app:zero2}
\end{equation}
Since the ratio between the amplitude of the particle-hole $f$ and the 
hole-particle $g$ components depends only on the energy and not on the 
particular state in the combination, the above equations ensure that the 
incoming flux of atoms and $R^-$ rotons vanish.  One has to recall the 
definition (\ref{eq:jj}) of the current density and use the values of $a_2$ 
and $a_3$ coming from Eqs.~(\ref{app:zero1}-\ref{app:zero2}) in order to  
calculate the flux of the incoming rotons and the ones of the outgoing 
excitations. The scattering rates come out to be:  
\begin{equation}
P_{++} = { \left| -i A_+^{(1)} e^{i q_+ b_1} +  
a_2 A_+^{(2)} e^{i q_+ b_2} + a_3 A_+^{(3)} e^{i q_+ b_3} \right|^2 \over 
\left|  i A_+^{(1)} e^{-i q_+ b_1} +  
a_2 A_+^{(2)} e^{-i q_+ b_2} + a_3 A_+^{(3)} e^{-i q_+ b_3}\right|^2  } 
\end{equation}
\begin{equation}
P_{+-} = {  \left|  i A_-^{(1)} e^{-i q_- b_1} +  
a_2 A_-^{(2)} e^{-i q_- b_2} + a_3 A_-^{(3)} e^{-i q_- b_3} \right|^2 \over 
\left|  i A_+^{(1)} e^{-i q_+ b_1} +  
a_2 A_+^{(2)} e^{-i q_+ b_2} + a_3 A_+^{(3)} e^{-i q_+ b_3}\right|^2  } 
\end{equation}
\begin{equation}
P_{+a} = { \left| A_a^{(1)} e^{-i q_a b_1} +  
a_2 A_a^{(2)} e^{-i q_a b_2} + a_3 A_a^{(3)} e^{-i q_a b_3}  \right|^2 \over 
\left|  i A_+^{(1)} e^{-i q_+ b_1} +  
a_2 A_+^{(2)} e^{-i q_+ b_2} + a_3 A_+^{(3)} e^{-i q_+ b_3}\right|^2  } 
\end{equation}
where $A_j=sign(f_j) \sqrt{v_j(f_j^2 - g_j^2)}$. One can also derive similar
expressions  using two antisymmetric states and one symmetric state.  The
other   scattering rates can be evaluated by selecting
different  processes, i.e., by solving different equations for the
coefficients $a_2$ and  $a_3$. The states entering the combinations have to
be linearly  independent. This requires a certain care in selecting proper
values of  $L$  and $b$.  If the states  are not sufficiently independent,
it happens that the  denominators in the expressions for $P_{ij}$ tend to be
small and  consequently small fluctuations in the numerical results produce
large  uncertainties in the final results. 

\section{Unitarity condition in the absence of $R^+$ normal reflection}
\label{app:d}

In this appendix we will derive expressions
(\ref{eq:special1}-\ref{eq:special4}) for the scattering matrix elements in
the phonon forbidden region (region III in the spectrum of
Fig.~\ref{fig:spectrum}) which are valid if $S_{++}$=0.  In this hypothesis
one can rewrite Eqs.~(\ref{eq:unitarity1}), (\ref{eq:unitarity*1}) and
(\ref{eq:unitarity*}), respectively, as 
\begin{equation}
|S_{+a}|^2 + |S_{+-}|^2 = 1
\label{eq:d1}
\end{equation}
\begin{equation}
S_{aa}^* S_{+a} + S_{a-}^* S_{+-} = 0 
\label{eq:d2}
\end{equation}
\begin{equation}
S_{+a}^* S_{-a} + S_{+-}^* S_{--}  = 0 \; .
\label{eq:d3}
\end{equation}
The first equation coincides with the result (\ref{eq:special1}). From the
other two one gets
\begin{equation}
S_{+-}^{*2} S_{--} = S_{+a}^{*2} S_{aa}
\end{equation} 
or
\begin{equation}
|S_{+-}|^4 |S_{--}|^2 = |S_{+a}|^4 |S_{aa}|^2 \; . 
\label{eq:d4}
\end{equation}
Subtracting Eq.~(\ref{eq:unitarity3}) from Eq.~(\ref{eq:unitarity2}) one
gets
\begin{equation}
|S_{+-}|^2 - |S_{+a}|^2 = |S_{aa}|^2 - |S_{--}|^2 \; .
\label{eq:d5}
\end{equation}
Excluding $S_{aa}$ from Eqs.~(\ref{eq:d4}) and (\ref{eq:d5}) one finds
\begin{equation}
|S_{+-}|^2 - |S_{+a}|^2 = \left( { |S_{-+}|^4 \over |S_{+a}|^4} -1 \right)
|S_{--}|^2
\end{equation}
or, using (\ref{eq:d1}): 
\begin{equation}
|S_{--}|^2 = \left( { |S_{+a}|^4 ( |S_{+-}|^2 - |S_{+a}|^2) \over
|S_{+-}|^4 - |S_{+a}|^4 } \right) = 
\left( { |S_{+a}|^4 \over |S_{+-}|^2 + |S_{+a}|^2 } \right)
= |S_{+a}|^4
\label{eq:d6}
\end{equation}
Thus we have obtained expression (\ref{eq:special2}). Expression
(\ref{eq:special3}) can be derived in analogous way. Finally, by
substituting (\ref{eq:d1}) and (\ref{eq:d6}) into (\ref{eq:unitarity2}), one
finds
\begin{eqnarray}
|S_{-a}|^2 &=& 1 - |S_{+-}|^2 - |S_{--}|^2 \nonumber \\
&=& |S_{+a}|^2 - |S_{+a}|^4 \nonumber \\
&=& |S_{+a}|^2 |S_{-+}|^2
\label{eq:d7}
\end{eqnarray}
which corresponds to the last result given in Eq.~(\ref{eq:special4}).

\begin{figure}
\caption{Schematic picture for the spectrum of elementary excitations in 
presence of a free surface. Solid line: phonon-roton dispersion for 
excitations propagating in the direction parallel to the surface ($q_z=0$).
Dashed line: threshold for atom evaporation. Excitations propagating at
different angles ($q_z\ne 0$) fill regions I-VI, as explained in the
text. } 
\label{fig:spectrum} 
\end{figure}

\begin{figure}
\caption{Spectrum of elementary excitations for a slab $80$\AA\ thick in 
a box $150$\AA\ wide, as obtained solving numerically the TDDF equations.
Two surface modes (ripplons) are also visible below the phonon-roton 
dispersion. For simplicity, only states symmetric with respect to the center
of the slab are shown.} 
\label{fig:spectrum2}
\end{figure}

\begin{figure}
\caption{Example of functions $f(z)$, in arbitrary units,   and $g(z)$,  
in the same units but multiplied by $5$, for a state at energy $10.2$ K and
parallel wave vector  $q_x=0.7$\AA$^{-1}$ in the same slab as in
Fig.~\protect \ref{fig:spectrum2}. The density profile of the slab is shown
as a dot-dashed line.  In part $b$, the Fourier transform of $f(z)$ is
shown; it corresponds to the convolution of two delta functions, at  the 
$R^+$ and $R^-$ roton wave vectors, with the  transform of the step function
used to restrict the Fourier  analysis within the slab.  The function
$g(q_z)$, not shown,  has a similar structure, but much smaller amplitude. 
} 
\label{fig:exampleofstate}   
\end{figure}

\begin{figure}
\caption{Schematic picture for  scattering processes at the free surface.
In the lower part a particular process is selected: an incoming $R^+$ roton
can produce atom evaporation, normal reflection or mode-change
reflection. } 
\label{fig:schemeofscatt} 
\end{figure}

\begin{figure}
\caption{Results for probabilities $P_{ij}$ as a function of energy, in the
phonon forbidden region.  } 
\label{fig:results}
\end{figure}

\begin{figure}
\caption{Ratio of the evaporation probabilities for $R^+$ and $R^-$ rotons
as a function of energy. Open circles: $q_x=0.6$\AA$^{-1}$; solid circles:
$q_x=0.65$\AA$^{-1}$; open squares: $q_x=0.7$\AA$^{-1}$; solid  squares:
$q_x=0.75$\AA$^{-1}$; triangles: $q_x=0.8$\AA$^{-1}$. Dashed line:
limiting value $P_{+a}/P_{-a}=1$ for energy approaching $\Delta=8.6$K. } 
\label{fig:p+/p-} 
\end{figure}

\begin{figure}
\caption{Schematic picture for the geometry used in Appendix C.  }
\label{fig:geometry}
\end{figure}

\begin{table}
\caption{Probabilities $P_{ij}$ for three different scattering processes
(incident atom, $R^+$ and $R^-$, respectively) described by linear 
combinations of four states at the same energy, $10.87$ K, and the same 
parallel wave vector, $0.7$\AA$^{-1}$. The states 1-4 are 
symmetric (2 and 4) and antisymmetric (1 and 3) solutions for
$L_{slab}=62$\AA\ (1 and 2) and $63.7$\AA (3 and 4), and for $L_{box}=
126$\AA\ (3), $134$\AA\ (4) and  $135$\AA \ (1 and 2).  The quantity
$\Sigma$ is the  sum of the probabilities for each process.  }  
\begin{tabular}{| r | r r r r | r r r r | r r r r |} 
\hline \multicolumn{1}{|c|}{}&
\multicolumn{4}{c|}{atom $\to$ atom, $R^+$, $R^-$}&
\multicolumn{4}{c|}{$R^+$ \ $\to$ atom, $R^+$, $R^-$}&
\multicolumn{4}{c |}{$R^-$ \ $\to$ atom, $R^+$, $R^-$}\\
\hline
\multicolumn{1}{|c|}{comb.}&
\multicolumn{1}{c}{$P_{aa}$}&
\multicolumn{1}{c}{$P_{a+}$}&
\multicolumn{1}{c}{$P_{a-}$}&
\multicolumn{1}{c|}{$\Sigma$}&
\multicolumn{1}{c}{$P_{++}$}&
\multicolumn{1}{c}{$P_{+-}$}&
\multicolumn{1}{c}{$P_{+a}$}&
\multicolumn{1}{c|}{$\Sigma$}&
\multicolumn{1}{c}{$P_{--}$}&
\multicolumn{1}{c}{$P_{-+}$}&
\multicolumn{1}{c}{$P_{-a}$}&
\multicolumn{1}{c |}{$\Sigma$}\\
\hline
\hline
1-2-3 & 0.148 & 0.654 & 0.224 & 1.026 & 0.0004 & 0.357 & 0.603 & 0.960 &
0.404 & 0.387 & 0.224 & 1.015 \\
1-2-4 & 0.146 & 0.645 & 0.224 & 1.015 & 0.0003 & 0.358 & 0.612 & 0.970 &
0.403 & 0.385 & 0.228 & 1.016 \\ 
1-3-4 & 0.148 & 0.650 & 0.223 & 1.021 & 0.0004 & 0.358 & 0.605 & 0.963 &
0.403 & 0.387 & 0.225 & 1.015 \\
3-2-4 & 0.146 & 0.645 & 0.224 & 1.015 & 0.0004 & 0.358 & 0.612 & 0.970 &
0.404 & 0.384 & 0.228 & 1.016 \\
\hline
\end{tabular}
\label{tab:combinations}
\end{table}

\begin{table}
\caption{ Values of the parameters used in $V_J({\bf r})$, 
see Eq.~(\protect \ref{app:ebf}). }

\begin{tabular}{c|c|c|c|c|c}                               
$\gamma_{11}$&$\gamma_{21}$&$\gamma_{12}$&$\gamma_{22}$&
$\alpha_1$&$\alpha_2$
\\
\tableline
$-19.7544$&$-0.2395$&$12.5616$ \AA$^{-2}$&$0.0312$ \AA$^{-2}$
&$1.023$ \AA$^{-2}$&$0.14912$ \AA$^{-2}$\\
\end{tabular}

\label{table}
\end{table}


\begin{references}
\bibitem{note1} Note that in the same expansion written in Ref.~\protect
\cite{Dal95} (Eq.(1)) there was a misprint in the sign of the argument of
the exponential after $g(z)$.
\bibitem{Bogoliubov} The functions $f$ and $g$ in the notation of standard
Bogoliubov theory are called $u$ and $v$, respectively.
\bibitem{note3} Different  states at the same energy and wave vector can be 
found by varying the thickness of the slab and/or the size of the 
computational box. This procedure is equivalent to solve the equations of 
motion at fixed energy, slab and box,  but different boundary conditions. 
We work with fixed boundary conditions ($f$ and $g$ vanish at the 
boundary).
\end{references}
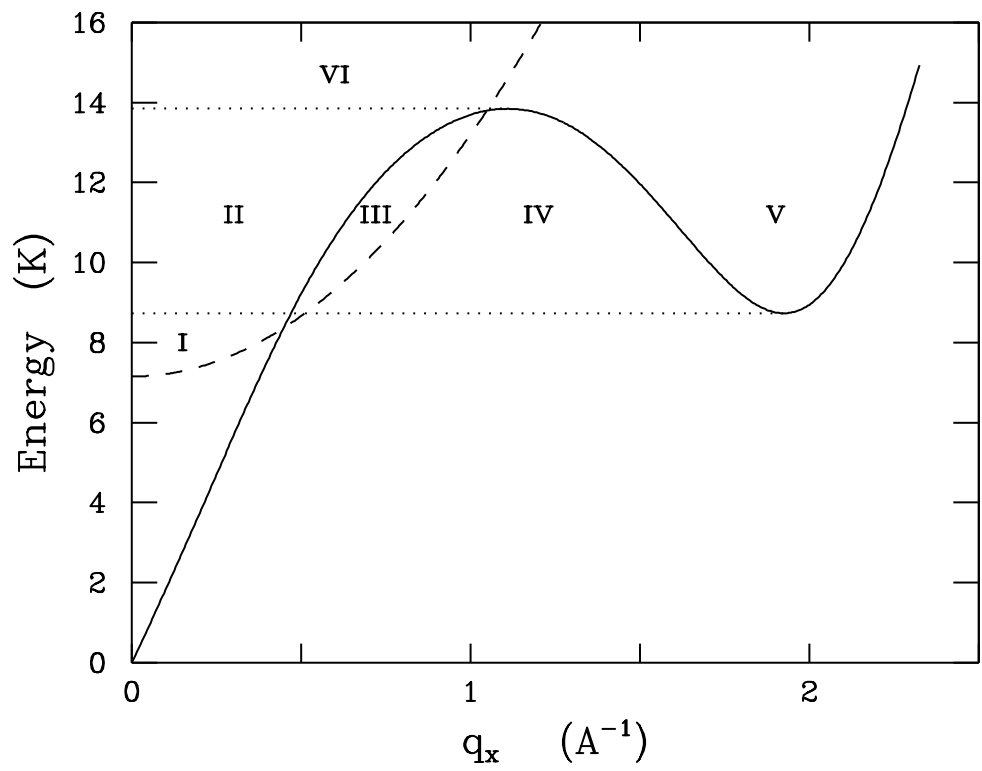
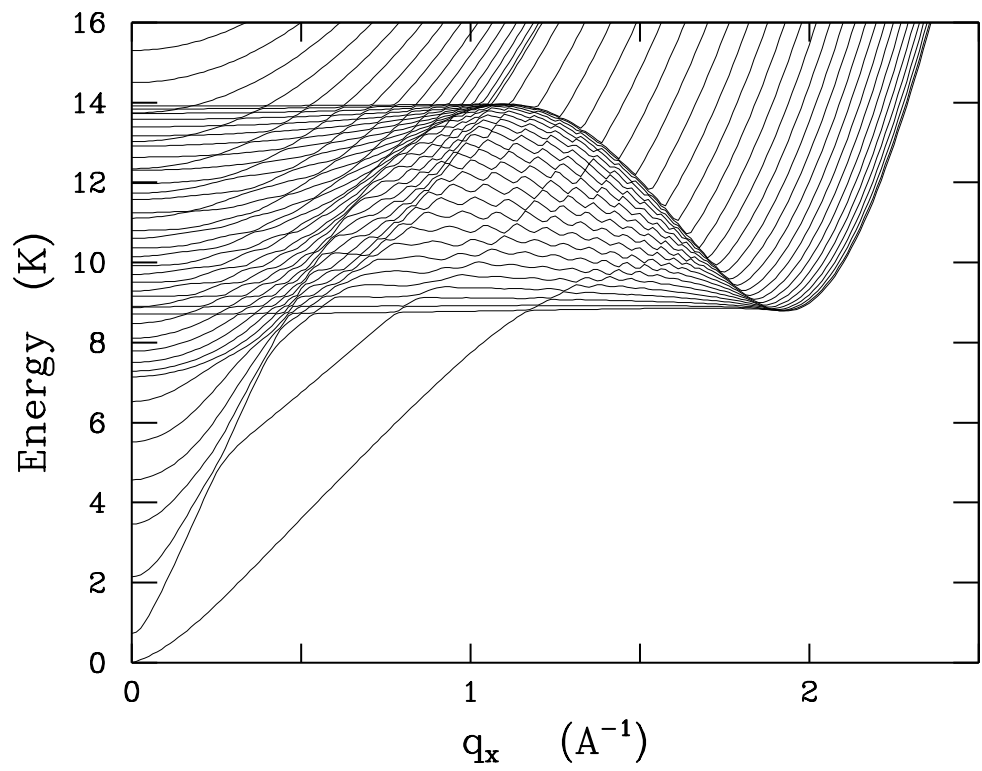
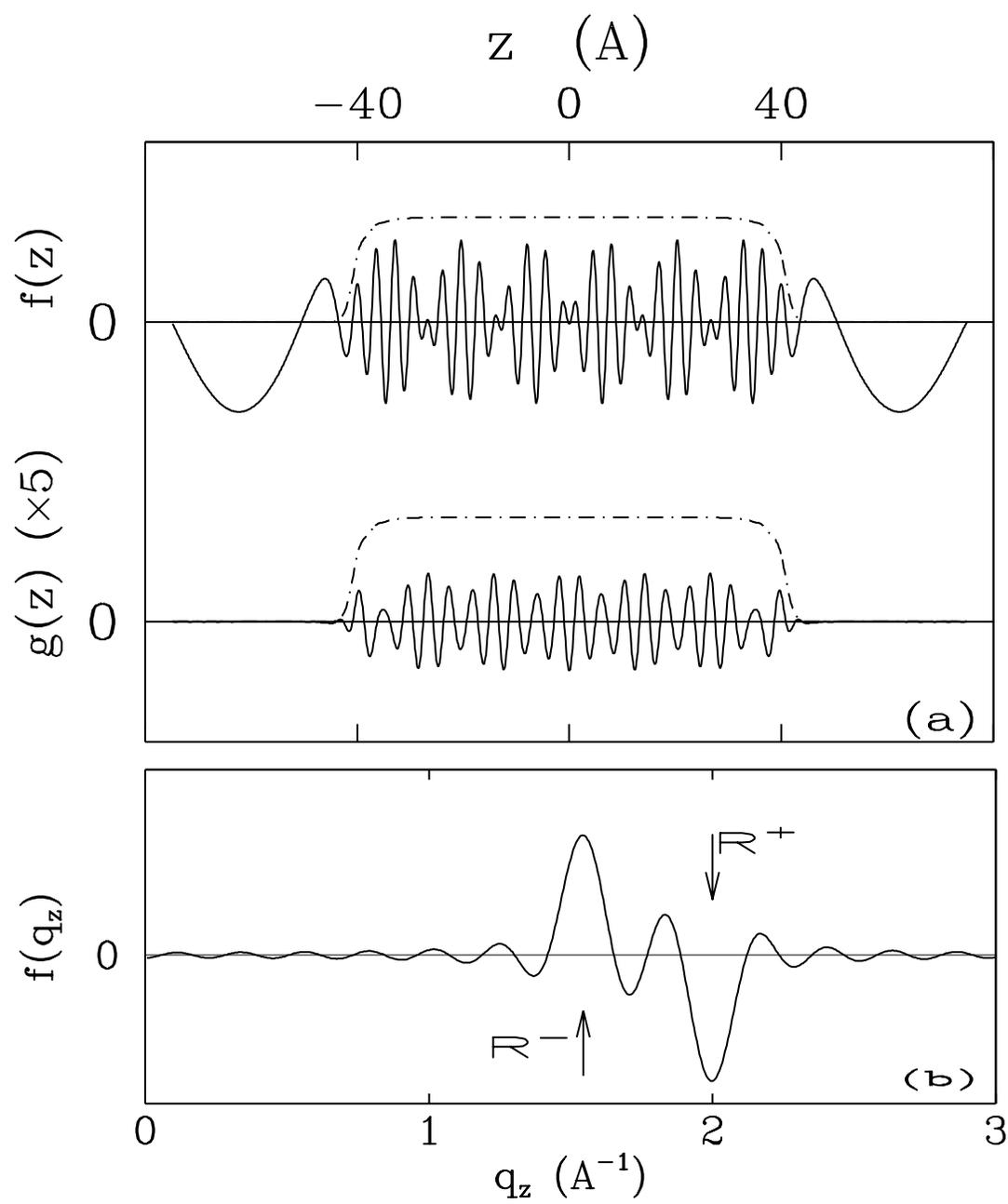
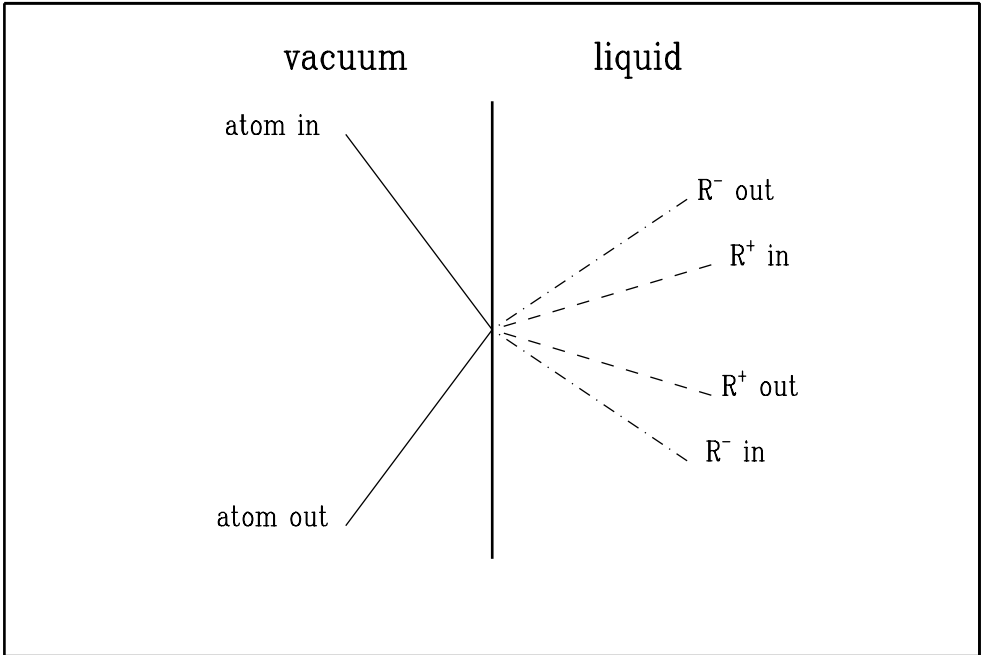
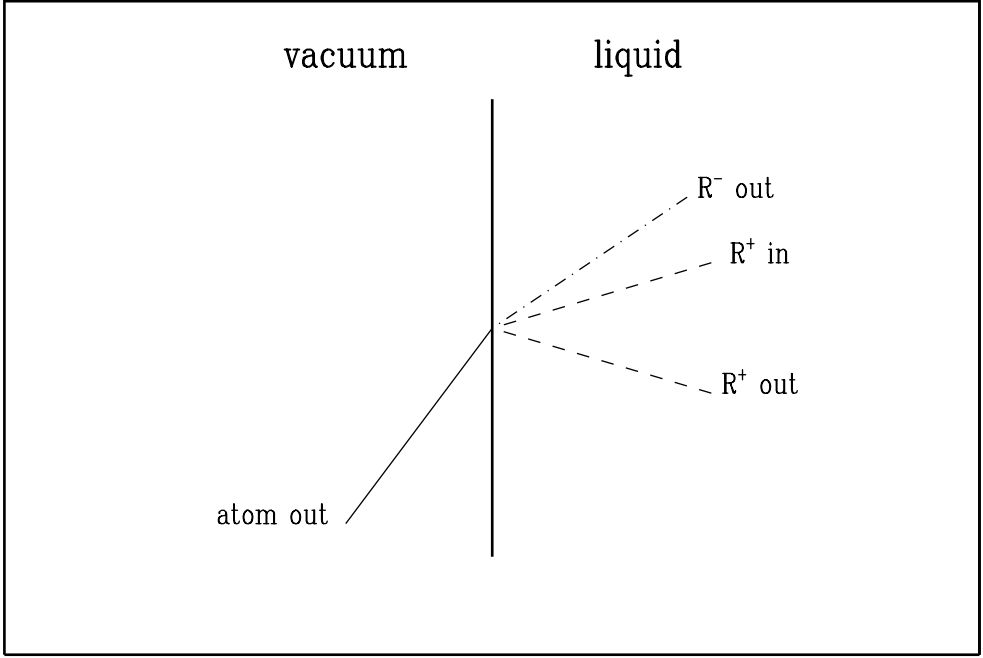
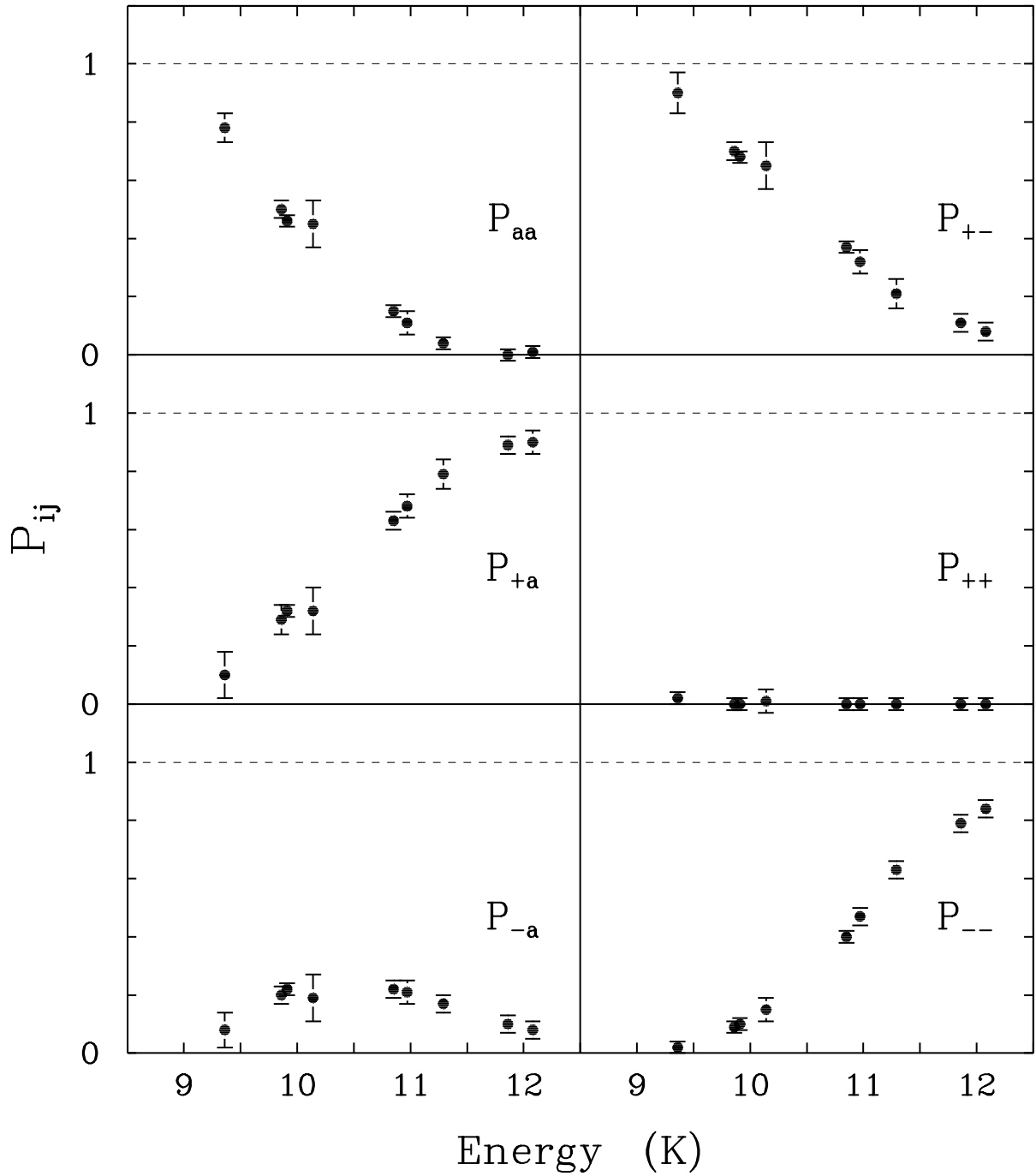
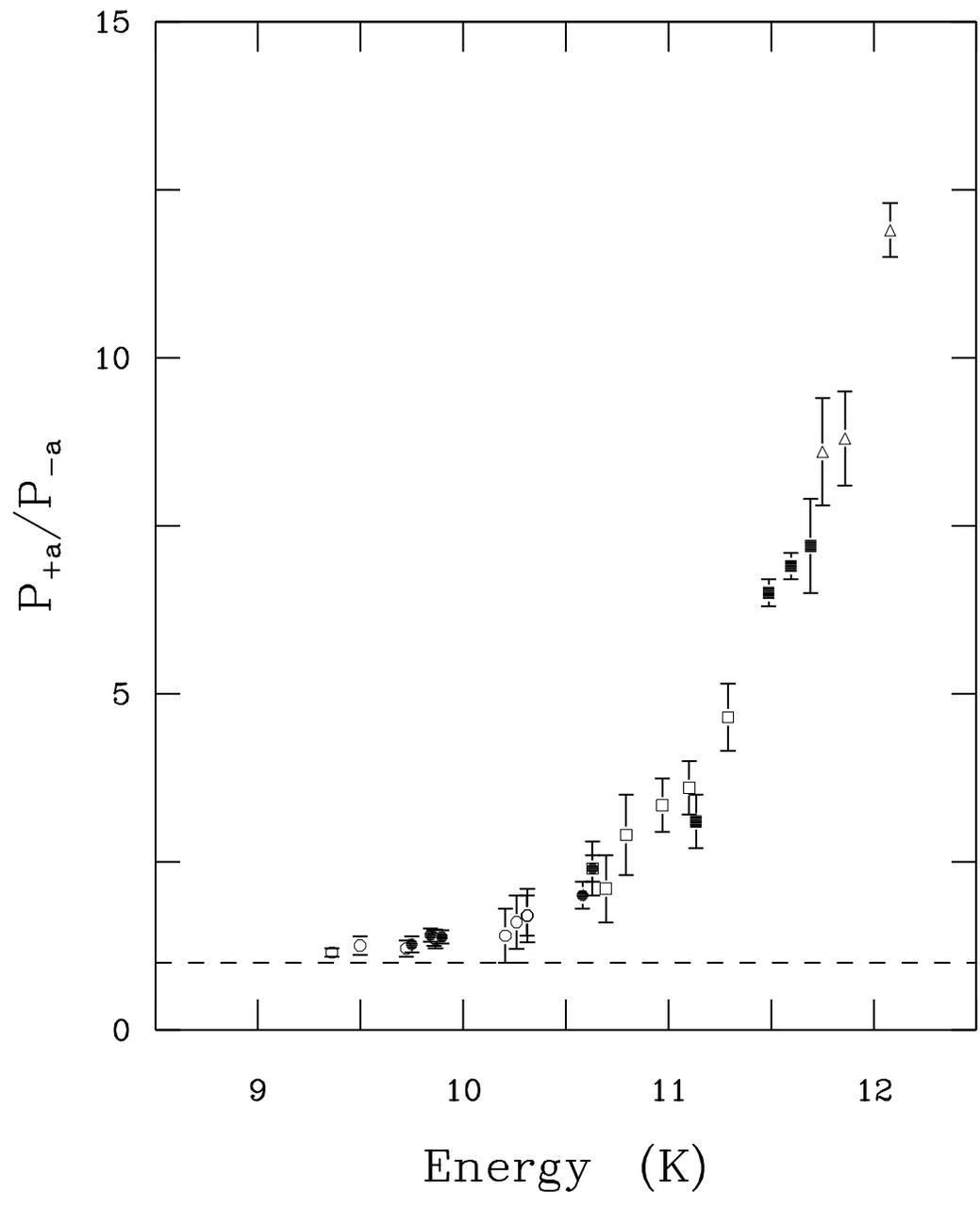
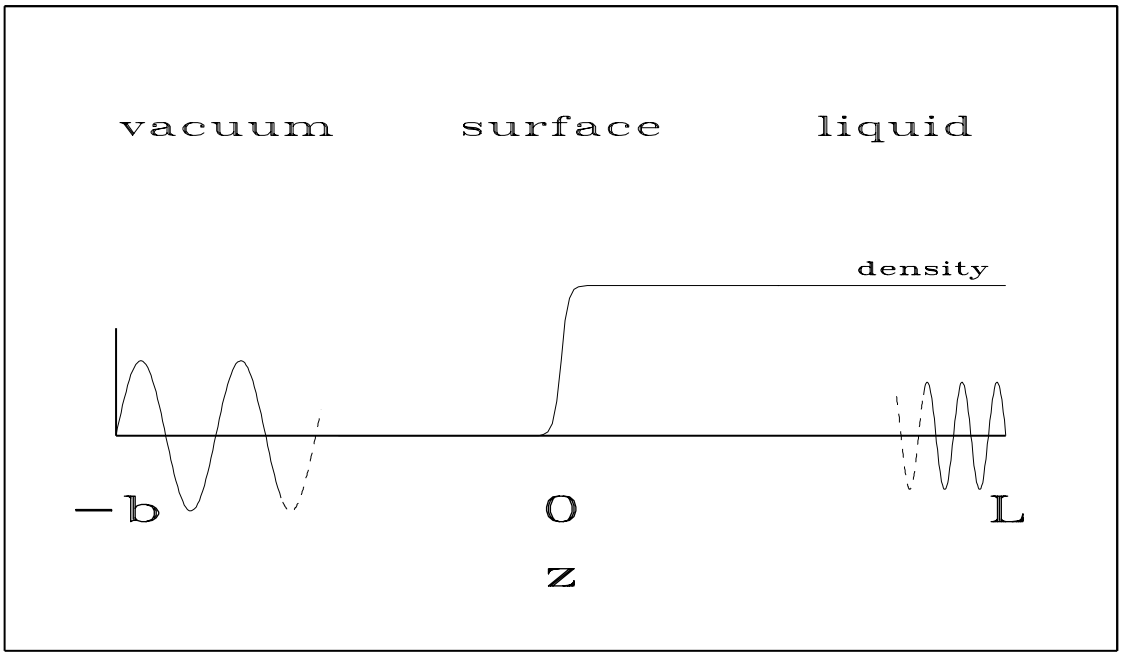
\end{document}